\documentstyle[pra,floats,aps,twocolumn,epsfig]{revtex}
\begin{document}\bibliographystyle{apsrev}

\wideabs{
\title{Mode-Locked Two-Photon States}
\author{Y. J. Lu, R. L. Campbell, and Z. Y. Ou}
\address{Department of Physics, Indiana
University-Purdue University Indianapolis \\ 402 N Blackford
Street, Indianapolis, In 46202}
\date{\today}
\maketitle

\widetext
\begin{abstract}
The concept of mode locking in laser is applied to a two-photon state with frequency entanglement.
Cavity enhanced parametric down-conversion is found to produce exactly such a state. The mode-locked
two-photon state exhibits a comb-like correlation function. An unbalanced Hong-Ou-Mandel
type interferometer is used to measure the correlation function. A revival of the typical interference
dip is observed. We will discuss schemes for engineering of quantum states in time domain.

\end{abstract}

}

\narrowtext
Parametric down-conversion process is known to produce two-photon state with entanglement in a variety of
degrees of freedom such as polarization\cite{shih}, phases\cite{grangier}, frequency\cite{om}, and
angular momentum\cite{Mair}. Because of its relative ease of production, polarization entanglement is 
mostly used in applications in quantum information\cite{mattle,bou}. More recently, attention has been
focussed on the spatial entanglement such as transverse modes\cite{walborn}. With new degrees of
entanglement discovered, there are more possibilities for information encoding. Among the entanglement
properties, seldom discussed is the temporal entanglement. This is not a  surprise if we consider the
fact that  the bandwidth ($\sim 10^9$ Hz) of current optical detectors cannot match that of the
down-conversion ($\sim 10^{12}$ Hz). Nevertheless, frequency (complementary to time) entanglement was
investigated recently for the potential nonlocal temporal shaping\cite{bellini}. A similar investigation
was done earlier by Zou et al\cite{zou}. Entanglement in the frequency domain involves infinite
dimensions of continuous Hilbert space and therefore should exhibit far richer physical phenomena.  In
this letter, we will study directly the temporal entanglement in a special situation similar to a
mode-locked laser and propose ways for quantum state engineering in the time domain by two-photon
interference.

The concept of mode locking was first introduced to produce short pulses from a laser\cite{siegman}.
Normally a free running laser emits optical fields in continuous waves (CW) which consist of many
independennt longitudinal modes of different frequencies. When the modes of the laser are locked in
phase, the output field becomes pulsed in a quasi-CW manner. The emitted pulses are spaced by the
cavity round trip time. The temporal behavior of the field is simply a reflection of the Fourier
transformation of the phase-locked frequency spectrum. Similarly, if the phases of different frequency
components of a two-photon state are locked, the result is a mode-locked two-photon state of the form: 
\begin{eqnarray}
|\Psi\rangle_{ML}  = \sum_{m = -N}^{N} \int &d\Omega&~ \psi (\Omega + m \Delta\Omega)\nonumber \\
\times &\hat a^{\dagger}& (\omega_p/2
+\Omega) \hat a^{\dagger} (\omega_p/2 - \Omega) |\rm{vac}\rangle, 
\end{eqnarray}
where $N$ is the number of frequency modes of correlated photons,
$\Delta\Omega$ is the frequency spacing between the adjacent modes, and $\psi(\Omega)$ gives the
spectral distribution for a single mode.  Different modes of photon pairs are in superposition, which
provides the mechanism for phase locking. Photons in each pair are correlated in
frequency.  Such a state can be generated from a parametric down-conversion filtered by a Fabry-Perot
cavity. The different frequency components come from the longitudinal modes of the cavity.
$\Delta\Omega$ is then the free spectral range of the cavity.  All the pairs have a common origin (phase)
from the pump field. The two-photon time correlation function can be calculated as 
\begin{eqnarray}
\Gamma^{(2)}(\tau)  &=& \langle \hat E^{(-)}(t)\hat E^{(-)}(t +\tau) \hat E^{(+)}(t +\tau)\hat
E^{(+)}(t)\rangle\nonumber\\ &=& \big |g(\tau) F(\tau)\big |^2, 
\end{eqnarray}
\begin{figure}[tbp]
\centerline{\epsfxsize=3.0in \epsffile{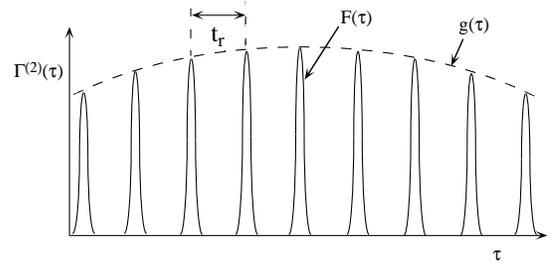}}
\caption{Comb-like time correlation function of a mode locked two-photon state in analogy to a mode
locked laser.}
\end{figure}
where 
\begin{eqnarray}
g(\tau) = \int d\Omega \psi(\Omega) e^{-i\Omega\tau},~
F(\tau) ={\sin[(2N+1)\Delta\Omega\tau/2]\over
\sin(\Delta\Omega\tau/2)}. \nonumber
\end{eqnarray}
Since $\psi(\Omega)$ is the spectrum of single mode, it has much narrower bandwidth than the full
bandwith $N\Delta\Omega$. So $g(\tau)$ is a slowly varying function and $\Gamma^{(2)}(\tau)$ is mainly
determined by the function $F(\tau)$, which has a comb-like shape (Fig.1). The period of $F(\tau)$ is
the cavity round trip time $t_r=1/\Delta \Omega$. The physics behind Eq.(2) is the following: when a pair
of photons enter the filter cavity, the cavity makes them bounce back and forth. Only when they hit the
output coupler, is there  some finite probability of escape and being detected. So the coincidence only
occurs at a time interval that is a multiple of the round trip time of the cavity. 

The comb-like time correlation function in Eq.(2) should be directly observable in a time delay
distribution measurement, provided that the resolution time $T_R$ of the detectors is smaller than the
time interval $t_r=1/\Delta\Omega$. Otherwise, the result is an average over the resolution time $T_R$
in many periods of $t_r$:
\begin{eqnarray}
\Gamma_{av}^{(2)}(\tau)  = A |g(\tau)|^2, 
\end{eqnarray}
where $A$ is a constant. So in the case of a poor detector resolution time, only the general contour of
$\Gamma^{(2)}(\tau)$ can be observed and the comb-like feature is lost. 

\begin{figure}[tbp]
\centerline{\epsfxsize=2.5in \epsffile{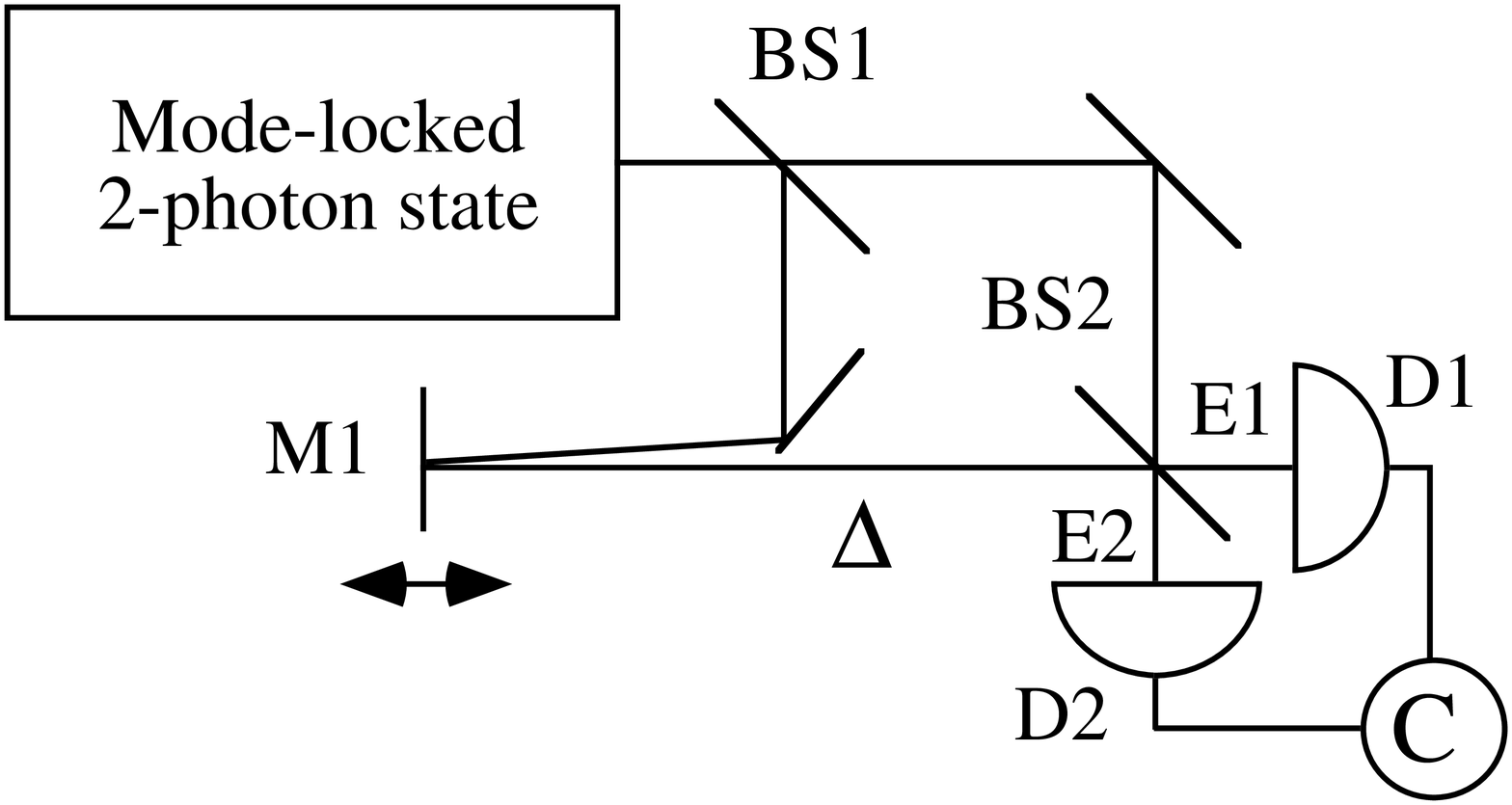}}
\caption{Layout of the interferometer. $\Delta$ is the time delay between the two arms.}
\end{figure}

However, the comb-like feature in Eq.(2) can be indirectly observed by the method of two-photon
interference with a variation of Hong-Ou-Mandel (HOM) interferometer\cite{hom} as shown in Fig.2. For a
collinear type-I parametric down-conversion, the two correlated photons co-propagate and can be separated
by a beam splitter (BS1). The second beam splitter (BS2) recombines the two photons to form the HOM
interferometer. The whole setup is just a Mach-Zehnder interferometer. With two-photon detection at the
outputs, it is also a Franson-type interferometer when the paths of the two arms are not
balanced\cite{fran,ozm}. In a simple single mode  model, the first beam splitter (BS1) transforms the
input two-photon state into the following state:
\begin{eqnarray}
|\Psi\rangle_{BS1} = (|2, 0\rangle + |0, 2\rangle +\sqrt{2}|1, 1\rangle)/2.
\end{eqnarray}
The first two terms give the usual two-photon interference (interference between short-short and
long-long paths) while the last term has no interference effect when the path difference is larger
than the coherence length and normally provides a constant background if the detectors cannot resolve
between the short and long paths. This will limit the maximum visibility to 50\%\cite{ozm}. With
mode-locked two-photon input, however, the comb-like correlation function indicates that the
$|1,1\rangle$ state will reappear at a path delay of every multiple of $ct_r$, the round trip distance of
the filter cavity. When this happens, the last term will exhibit Hong-Ou-Mandel interference 
dip\cite{hom} at nonzero delays. The revival of HOM interference dips was first predicted by
Shapiro\cite{shapiro}.  

The intuitive argument above can be easily confirmed by a calculation of the two-photon coincidence rate
between the two detectors at the output of the
unbalanced Mach-Zehnder interferometer in Fig.2. We use a multimode state given in Eq.(1) as the
input state to the interferometer and obtain the result with 50:50 beam splitters as follows:
\begin{eqnarray}
\Gamma_{12}^{(2)}(\tau)  &=& \langle \hat E_1^{(-)}(t)\hat E_2^{(-)}(t +\tau) \hat E_2^{(+)}(t +\tau)\hat
E_1^{(+)}(t)\rangle\nonumber\\ &=& {1\over 2}\big |g(\tau) F(\tau)\big |^2 (1 - \cos\omega_p\Delta)+
\nonumber\\&~&~+
{1\over 4}\big |g(\tau+\Delta) F(\tau+\Delta) - g(\tau - \Delta) F(\tau - \Delta)\big |^2 \nonumber \\
&~&~+ Re \big\{ i\sin(\omega_p\Delta/2) g(\tau)F(\tau) [g(\tau+\Delta)\times\nonumber\\
&~&~\times  F(\tau+\Delta) - g(\tau - \Delta) F(\tau - \Delta)] \big\}. 
\end{eqnarray}
The last term gives no contribution when it is
integrated over the detector's resolving time $T_R$ that is larger than the time delay $\Delta$. So the
two-photon coincidence rate is proportional to
\begin{eqnarray}
R_2(\Delta) &=& \int_{T_R} d\tau \Gamma_{12}^{(2)}(\tau)\nonumber\\  &=&  {R_0\over 2}  (1 -
\cos\omega_p\Delta) + {R_0\over 2} [1- V(\Delta)] 
\end{eqnarray}
where
\begin{eqnarray}
R_0&=& \int_{T_R} d\tau \big |g(\tau) F(\tau)\big |^2 
\nonumber\\
V(\Delta)&=& {\int_{T_R} d\tau ~g(\tau+\Delta) F(\tau+\Delta) g(\tau - \Delta) F(\tau - \Delta)\over
\int_{T_R} d\tau \big |g(\tau) F(\tau)\big |^2}. \nonumber 
\end{eqnarray}
The first term in Eq.(6) corresponds to the first two terms in Eq.(4)
and produces a phase sensitive two-photon interference pattern. The second term in Eq.(6) arises from the
last term in Eq.(4) and gives rise to the HOM interference dip as $\Delta$
is scanned. Normally, there is only one dip around zero delay ($\Delta \approx 0$). But for mode-locked
two-photon state, the reappearance of the coincidence peak at nonzero delays (due to the comb-like
correlation function) will revive the HOM dips every time when the time delay $\Delta $ is such that
$F(\tau+\Delta)$ overlaps with
$F(\tau-\Delta)$. This corresponds to $\Delta = M t_r/2$ with $M =$ interger. A surprising result is that
the period of the revival of HOM dip is $t_r/2$ rather than $t_r$ predicted from a previous simple
intuitive argument and Ref.\cite{shapiro}. The shorter period can be
understood if we take a detailed look at the timeline of photodetection in Fig.3 of an unbalanced HOM
interferometer. The figure shows the interference of two possibilities: both photons are transmitted or
both are reflected. In each case,
$2(l_1-l_2)$ is the path difference between the two arms of the interferometer. Fig.3a corresponds to the
intuitive argument: the two photons come from adjacent coincidence peaks with $\Delta = t_r$. In Fig.3b,
photodetections of the two photons are not simultaneous but have a time difference of
$t_c/2$\cite{pittman}. The two overlapping possibilities are from two different cases: two photons are
separated by a delay of $t_r$ or they are simultaneous. Because of mode lock nature of the process, the
two possibilities are coherent to each other and will produce interference. In this case, we only need a
time delay $\Delta$ to be $t_r/2$.
\begin{figure}[tbp]
\centerline{\epsfxsize=2.5in \epsffile{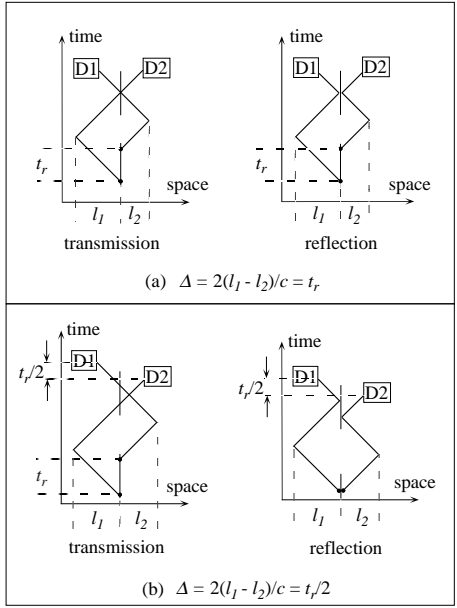}}
\caption{Timeline for photodetection of two photons in an unbalanced HOM interferometer. See
text for details.}
\end{figure} 

Although filtering after the generation of parametric down-converted photons will produce the required
mode-locked two-photon state, it is at the expense of signal level, for the
down-converted light signal is proportional to the detection bandwidth.  Recently we have successfully
implemented a type-I optical parametric oscillator (OPO) far below threshold for the generation of narrow
band two-photon state without the reduction
of the signal level\cite{ou}. Multi-mode operation of the device  produces naturally a mode locked
two-photon state.  The cavity round trip time of the device is of the order of 1 ps, which prevents us
from direct observation of the comb-like correlation function in Eq.(2).  Nevertheless, we did observe
the time average behavior predicted in Eq.(3) in a time delay distribution measurement. 

\begin{figure}[tbp]
\centerline{\epsfxsize=2.7in \epsffile{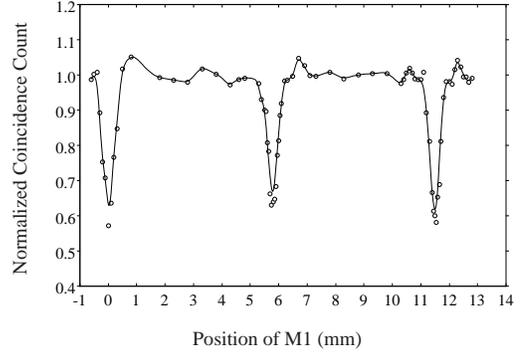}}
\caption{Normalized coincidence as a function of the micrometer position of mirror M1. The solid line is a
smooth interpolation of the data for visual guidance.}
\end{figure}

To indirectly show the mode locking effect, we input the state into an unbalanced Mach-Zehnder
interferometer as sketched in Fig.2 and observe the coincidence count between the two outputs as 
the mirror M1 is scanned. The mirror M1 is mounted on a piezo-electric transducer for phase scan and a
micrometer for large range location scan. The coincidence window is measured to be 10ns. Under this
condition ($T_R = 10$ ns $>>
\Delta$), The coincidence rate is given by Eq.(6). The first term of Eq.(6) is a phase
dependent term that is always there. In order to concentrate on the second term in Eq.(6) for unbalanced
HOM interference effect, we dither the phase (piezoelectricc transducer) so that the
contribution from the first term is merely a constant baseline that will limit the HOM interference
visibility to a maximum of 50\%. In Fig.4, we plot the corrected coincidence counts as a function of
the position of M1 (micrometer).  The reappearance of the HOM dip at nonzero delays in Fig.4 implies a
two-photon correlation function as in Eq.(2)\cite{shapiro}.  The data was collected in separate
experiments because the interferometer needs to be realigned after some large displacement of M1 (the
visibility of the interferometer, which is independently monitored by an auxiliary laser,  drops
significantly after about 6 mm displacement of M1). So the coincidence data has to be normalized to an
average of the points at the wings of the dips. The spacing between dips is 5.75 mm corresponding to one
half cavity round trip distance of the OPO cavity. 
\begin{figure}[tbp]
\centerline{\epsfxsize=3.0in \epsffile{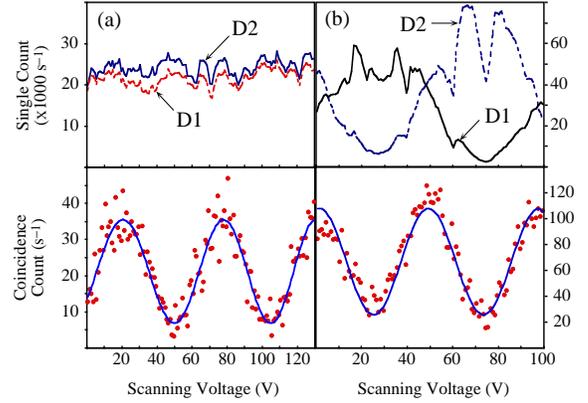}}
\caption{Coincidence as well as single counts as a function of the voltage of piezoelectric transducer.
Micrometer for M1 is set at (a) 5.7 mm and (b) 11.5 mm. }
\end{figure}

Next we fix the micrometer position of M1 at the bottom of the two dips with nonzero path differences
which correspond to one half and one full cavity round trip distance, respectively. We then scan the
phase via the piezoelectric transducer. Fig.5 shows the coincidence as well as the single detector counts
as a function of electric voltage at the two micrometer positions of M1. 
Coincidence counts at both positions show the
sinusoidal interference pattern with visibilities larger than 50\%. The solid curves is a least square
fit to a SINE function with 68\% and 62\% visibility, respectively. The low visibility is attributed to
poor mode match at large path delays. A surpise from Fig.5b shows that the single
detector counts also vary sinusoidally with the phase change and the counts from the two detectors are
180 degree out of phase (The unexpected drops in single counts are due to instability of the OPO cavity
and are corrected in coincidence counts). So the interference pattern in coincidence is simply from the
anti-correlation of single counts. This is not fourth-order but second-order interference. The
reappearance of second-order coherence at nonzero delay can be easily understood by calculating the
second-order field correlation function:
\begin{eqnarray}
\gamma(\tau)  &=& \langle\Psi| \hat E^{(-)}(t+\tau) \hat E^{(+)}(t)|\Psi\rangle\nonumber\\ &=& 
e^{i\omega_p\tau/2} G(\tau) F(\tau)
\end{eqnarray}
with $G(\tau) = \int d\omega |\psi(\omega)|^2 e^{i\omega\tau}$. $|\gamma(\Delta)|$ gives the
visibility of interference patterns in single detector counts and it has similar comb-like shape as
$\Gamma^{(2)}(\tau)$. So the single count interference pattern revives at various multiples of $t_r$, 
just like a mode locked laser. In contrast, interference pattern in coincidence occurs with a period of
$t_r/2$. For those micrometer positions of M1 that are not inside any of the dips in Fig.4, no
interference arises from the second term of Eq.(6). This term simply adds a constant to the baseline to
reduce the vibility to maximum of 50\%.  This corresponds
to the simple scheme of Franson interferometer\cite{fran}. We observed a visibilty of around 35\% at those
locations.

The interesting comb-like correlation function can be used for quantum state engineering. Here we
propose to use two-photon interference to take out one of the spikes in the correlation function (Fig.1).
To do that, we consider a wide band two-photon state described by 
\begin{eqnarray}
|\Phi\rangle_{WB}  = \int &d\Omega&~\phi (\Omega)~ e^{i(\omega_p/2 - \Omega)\delta t}\nonumber \\
\times &\hat b^{\dagger}& (\omega_p/2
+\Omega) \hat b^{\dagger} (\omega_p/2 - \Omega) |\rm{vac}\rangle, 
\end{eqnarray}
where $\phi(\Omega)$ gives the wide spectrum of down-conversion and $\delta t$ sets a relative delay
betwen the two photons. The two-photon correlation function is simply
\begin{eqnarray}
\Gamma^{(2)}(\tau)  = \big |f(\tau - \delta t) |^2 ~~
{\rm with}~~ 
f(\tau) = \int d\Omega \phi(\Omega) e^{-i\Omega\tau}. \nonumber
\end{eqnarray}
This is a single peaked function centered at $\delta t$. 

We mix this state with the mode-locked two-photon state in Eq.(1). The actual state of
the system is 
\begin{eqnarray}
|\chi\rangle  = (|\rm{vac}\rangle +\eta|\Psi\rangle_{ML})\otimes (|\rm{vac}\rangle
+\zeta|\Phi\rangle_{WB}). 
\end{eqnarray}
Here we add in the vacuum state to write the true states from parametric down-conversion and the
coefficients $\eta$ and $\zeta$ are related to a common pump field. We can easily calculate the time
correlation function of the combined field as
\begin{eqnarray}
\Gamma^{(2)}(\tau)  = \big |\eta g(\tau)F(\tau) + \zeta f(\tau - \delta t) \big |^2 
\end{eqnarray}
If $f(\tau - \delta t)$ overlaps with one of the peaks of $F(\tau)$, destructive interference will take
out that peak with proper adjustment of $\eta$ and $\zeta$. By changing the delay $\delta t$, we can
manage to take any one out for information coding. 

In conclusion, we have applied the concept of mode locking to entangled two-photon state and observed its
effects in an unbalanced HOM interferometer. Quantum interference can be used to manipulate the
entanglement in time domain.

\acknowledgements
We'd like to thank P. Kumar for pointing us to Ref.\cite{shapiro} and suggesting a possible experiment. 
This work was supported by Purdue Research Foundation and NSF.

\end{document}